\documentclass[aps,prx,twocolumn,groupedaddress,nofootinbib,superscriptaddress,showpacs]{revtex4-2}

\usepackage{hyperref}       
\usepackage{url}            
\usepackage{booktabs}       
\usepackage{amsfonts}       
\usepackage{amssymb}
\usepackage{mathtools}
\usepackage{nicefrac}       
\usepackage{microtype}      
\usepackage{amsmath}
\usepackage{xspace}
\usepackage{dsfont}
\usepackage{cases}
\usepackage{pifont}
\usepackage[export]{adjustbox}
\usepackage{relsize}
\usepackage{xcolor,colortbl}
\usepackage{hhline}
\usepackage{makecell}
\usepackage{physics}
\usepackage{bbm}
\usepackage{changes}
\usepackage{tabularx}
\usepackage{amsthm}
\usepackage[nameinlink]{cleveref}
\usepackage{changes}
\usepackage{tikz}
\usepackage{color,soul}
\usepackage{multirow}
\usepackage{braket}

\def\ie{\textit{i.e.}}
\def\eg{\textit{e.g.}}
\newcommand{\gsim}{\raisebox{-0.13cm}{~\shortstack{$>$ \\[-0.07cm]
      $\sim$}}~}

  \definecolor{masoncolor}{rgb}{0.98, 0.27, 0.62}

\begin{document}

\title{Modeling and forecasting age-specific drug overdose mortality in the United States}
%
%
\author{Lucas B\"{o}ttcher}
\email{l.boettcher@fs.de}
\affiliation{Dept.~of Computational Science and Philosophy, Frankfurt School of Finance and Management, Frankfurt am Main, 60322, Germany}
\author{Tom Chou}
\affiliation{Dept.~of Computational Medicine, UCLA, Los Angeles, CA 90095-1766}
\affiliation{Dept.~of Mathematics, UCLA, Los Angeles, CA 90095-1555}
\author{Maria R. D'Orsogna}
\affiliation{Dept.~of Mathematics, California State University at Northridge, Los Angeles, CA 91330-8313}
\affiliation{Dept.~of Computational Medicine, UCLA, Los Angeles, CA 90095-1766}
\date{\today}
%

\begin{abstract}
Drug overdose deaths continue to increase in the United
  States for all major drug categories. Over the past two decades the
  total number of overdose fatalities has increased more than
  five-fold; since 2013 the surge in overdose rates is primarily
  driven by fentanyl and methamphetamines. Different drug categories
  and factors such as age, gender, and ethnicity are associated with
  different overdose mortality characteristics that may
    also change in time. For example, the average age at death from a
  drug overdose has decreased from 1940 to 1990 while the overall
  mortality rate has steadily increased. To provide insight into the
  population-level dynamics of drug-overdose mortality, we develop an
  age-structured model for drug addiction. Using an augmented ensemble
  Kalman filter (EnKF), we show through a simple example how our model
  can be combined with synthetic observation data to estimate
  mortality rate and an age-distribution parameter. Finally, we use an
  EnKF to combine our model with observation data on overdose
  fatalities in the United States from 1999 to 2020 to forecast the
  evolution of overdose trends and estimate model parameters.
\end{abstract}


\maketitle
\section{Introduction}\label{sec:intro}
The number of drug-overdose fatalities in the United States has been
steadily increasing over the past 20
years~\cite{jalal2018changing,JAL2020}. Between 1999 and 2020 more
than 900,000 drug-overdose deaths were reported in the US. In 2020
alone, almost 100,000 people died from injury or poisoning from drugs
of abuse (mainly opioids and psychostimulants), constituting a 32\%
rise over 2019. According to provisional mortality
data~\cite{ahmad2021drug}, this trend has continued throughout
2021.

A study~\cite{jalal2018changing} that examined the exponential growth
in drug overdose deaths between 1979 and 2016 in the US
reveals that the drug types causing these rises have changed over
time. During the 1980s and 1990s, the majority of fatal
  drug overdoses were due to illegal substances such as heroin and
  cocaine. Successive overdose waves were driven by prescription
  opioids in the 2000s, followed briefly by heroin in 2010, and,
  beginning in 2013, by synthetic opioids. The synthetic opioid wave
persists to this day as the majority of US overdose deaths are due to
fentanyl and its derivatives.  There is also substantial variability
in the demographic patterns of drug-overdose deaths. While cocaine and
prescription drugs mostly led to increased mortality among
40-to-50-year-olds, current fentanyl use is accompanied by fatality
rate increases among 20-to-40-olds. In addition to age, factors such
as gender, race, and place of residence are also associated with
variations in drug-overdose risk~\cite{d2023fentanyl}.
 
The majority of studies analyzing the spatio-temporal evolution of
overdose mortality are mainly descriptive and rely on data
visualization and statistical analysis of past data. In this work, we
instead use an age-structured
model~\cite{m1925applications,chou2016hierarchical} to mechanistically
study the drug epidemic in the US. The model is then used in
conjunction with empirical data to forecast the short-term evolution of overdose mortality through an
ensemble Kalman filter (EnKF), a data assimilation
technique~\cite{evensen1994sequential,crassidis2004optimal,brown2012introduction}.

Age-structured models (also known as Kermack--McKendrick models) can
be used to mathematically describe the evolution of distinct
population categories (\eg, susceptible and dead), where the dynamics
and interactions among categories may depend on the distribution of
age in the population. Different variants of age-structured models
have been developed and applied to model heroin addiction as an
epidemic~\cite{fang2015global,fang2015global2,YAN2016,LIU2019,CHE2020,DIN2020,DUA2021,KHA2021}. Such
models have also been applied to mechanistically describe cellular
processes~\cite{belair1995age,mahaffy1998hematopoietic} and population
dynamics associated with social interactions~\cite{chuang2018age}, birth control policies \cite{CHILDPOLICY}, and
COVID-19
mortality~\cite{bottcher2020case,richard2021age,kreck2022back}.

The ensemble Kalman filter, which we use to combine observation data
with an age-structured model of overdose mortality, originated from
research activities in the geophysical sciences and has found various
applications in problems that require combining high-dimensional
dynamical systems with observation
data~\cite{katzfuss2016understanding}. Kalman filtering and related
data assimilation methods (\eg, Bayesian Markov chain Monte Carlo)
have been used in computational biology and medicine to estimate model
parameters~\cite{lillacci2010parameter,pandey2013comparing,bomfim2020predicting,reis2020spatio},
identify patients with antibiotic-resistant bacteria in hospital
wards~\cite{pei2021identifying}, and develop risk-dependent contact
interventions in epidemic
management~\cite{schneider2021epidemic}. Within computational social
science, data assimilation methods have proven useful in combining
mechanistic models with survey data, \eg, to study the evolution of
political polarization in the US~\cite{bottcher2020great}.

In Sec.~\ref{sec:model} we present a general age-structured model that
includes age-dependent addiction and age-specific
mortality. We also discuss approximations that admit analytical
solutions. The basic concepts underlying the EnKF are outlined in
Sec.~\ref{sec:enkf}. In Sec.~\ref{sec:application_wonder} we adapt our
general age-structured model to describe a population suffering from
substance use disorder (SUD). We then describe the available
drug-overdose data and illustrate how the EnKF is applied to our model
and dataset. Finally, we conclude our work with a discussion and
future outlook in Sec.~\ref{sec:discussion}.
\section{A general age-structured mortality model}\label{sec:model}
Our starting point is the general age-structured model
\begin{align}
\begin{split}
\left[\frac {\partial} {\partial a}  + \frac {\partial} {\partial t} \right] 
n (a,  t)   &= 
- \mu (a, t) n (a, t)+ p(a,t),
\label{eq:pde_model}
\end{split}
\end{align}
where $n(a,t) \mathrm{d}a$ is the number of individuals (
  \textit{i.e.}, people with SUD in our application) with age between $a$ and
  $a + \mathrm{d}a$ at time $t$.  We assume this population dies at
  rate $\mu(a,t)$, and that there is an influx rate
  $p(a,t)$. The initial conditions at $t=t_0$ and $a=0$ are specified
  via $n(a, t=t_0) = \rho(a)$ and $n(a=0, t) = g(t)$, where $\rho$ and
  $g$ are non-negative functions such that $g(t \to t_0) = \rho(a \to
  0)$. We specifically set $g(t) = 0$, implying that no population of
  age $a=0$ exists at any time. In the context of
    modeling overdose mortality, this means that the number of
    addicted newborns is assumed to be zero. Note that this model is
  different from the original McKendrick model~\cite{m1925applications} in which an age-dependent birth
  rate generates newborns through a self-consistent boundary condition
  on $n(a,t)$.  To solve Eq.~\eqref{eq:pde_model} analytically we use
  the method of characteristics and distinguish the two cases $a \geq
  t - t_0$ and $a < t - t_0$.  For $a \geq t-t_0$, the characteristic
  will begin at $t=t_0$ and $n(a,t)$ will remain constant along $a = t
  -t_0$.  When $a < t - t_0$ the characteristic will begin at $a = 0$
  and $n(a,t)$ will remain constant along $t = a + t_0$. The formal
  solution to Eq.~\eqref{eq:pde_model} can then be expressed as

\begin{widetext}
\begin{numcases}{n(a,  t) =}
\rho (a -t + t_0) e^{-\int_{t_0}^t \mu(a-t+s, s)\, \mathrm{d}s}
+ \!\int_{t_0}^{t}\! p(s + a - t, s)e^{-\int_{s}^t \mu(a-t+s', s') \mathrm{d}s'} \mathrm{d}s &  $(a \geq t - t_0)$ 
\label{analysol1} \\
\int_{0}^{a}\! p(s, s +t -a )e^{-\int_{s}^a \mu(s', s' + t - a)\, \mathrm{d}s'} \mathrm{d}s  & $(a < t - t_0)$.
\label{analysol2}
\end{numcases}
As a specific example we set the initial time $t_0 = 0$, fix the
initial condition $\rho(a) = 0$, and impose a constant death rate
$\mu(a,t) = \mu$. We further assume an immigration rate $p(a,t) = p(a)
= a e^{-\lambda a}$ which has a maximum at age
$\lambda^{-1}>0$. Equations~\eqref{analysol1} and \eqref{analysol2} become

\begin{numcases}{n(a,  t) =}
{e^{-\lambda (a-t)} \over (\lambda - \mu)^2}
\Big[e^{-\mu t }\big(1 +  (a-t)(\lambda - \mu) \big) -  \big(1+ a(\lambda - \mu)\big) e^{-\lambda t}\Big] & $(a \geq t )$ \label{back} \\
\frac{1}{(\lambda - \mu)^2}\Big[e^{-\mu a} -\big(1 + a (\lambda - \mu)\big) e^{-\lambda a}\Big] & $(a < t)$. \label{front}
\end{numcases}
\end{widetext}

The function $p(a) = a e^{-\lambda a}$ describes an
  influx of people of mean age $2\lambda^{-1}$ that suffer from an
  SUD. Using this functional form, the number of SUD cases that are
  much younger/older than $2\lambda^{-1}$ is small compared to the
  number of SUD cases with an age of about $2\lambda^{-1}$. The
  distribution of overdose cases in the US population follows a
  qualitatively similar trend~\cite{abuse2019mental}. We use this
  analytically tractable example in Sec.~\ref{sec:enkf} to explain how
  age-structured models of the form presented in
  Eq.~\eqref{eq:pde_model} can be combined with Kalman filters to
  learn model parameters from noisy observations. In
  Sec.~\ref{sec:application_wonder}, we describe $p(a)$ by a more
  general linear combination of two gamma distributions to connect our
  model of drug-overdose mortality with corresponding data from the
  CDC WONDER database.

Observe that $n(a,t)$ in Eqs.~\eqref{back} and
  \eqref{front} is continuous for $a=t$ and that the maxima of
  Eq.~\eqref{back} are located along the trajectory
\begin{align}
\begin{split}
a_{\rm max} (t) = \frac{t}{1 - e^{-(\lambda - \mu) t}} - \frac {\mu}{\lambda (\lambda - \mu)}, 
\end{split}
\label{max}
\end{align} 
where $a_{\rm max}(t) > t$ is an increasing function of time.  The
steady state form of $n(a,t\to \infty)$ is given by the
time-independent term in Eq.~\eqref{front}.
\section{Ensemble Kalman filter}\label{sec:enkf}
In the first part of this section we describe the basic definitions
and update rules in the EnKF~\cite{evensen1994sequential}. We use the
standard state-space representation of a physical system and
distinguish between state, output, and input (\ie, control)
variables. Outputs are quantities that can be observed or measured
(\eg, the number of overdose deaths), while other quantities such as
age-specific mortality rates and the number of individuals suffering
from SUD are state variables that are not known and have to be
estimated.  As a first application example, we use the EnKF to
estimate the rates $\mu$ and $\lambda$ that arise in
Eqs.\,\eqref{back} and \eqref{front} of the simple model presented in
Sec.~\ref{sec:model}.

\subsection{Basic definitions}
To outline the main steps associated with the application of an EnKF
to the age-structured PDE model in Eq.\,\eqref{eq:pde_model}, we
primarily follow the notation of
Refs.~\cite{brown2012introduction,crassidis2004optimal}; the EnKF
implementation that we use in this work is instead based on
Ref.~\cite{labbe2014}.

The evolution of the system state $\mathbf{x}(t)$ and observed state
$\mathbf{z}(t)$ is described by
\begin{align}
\begin{split}
\dot{\mathbf{x}}&=\mathbf{f}(\mathbf{x},t)+\mathbf{w}(t)\quad\mathbf{w}(t)\sim\mathcal{N}(\mathbf{0},\mathbf{Q}(t))\\
\mathbf{z}&=\mathbf{h}(\mathbf{x},t)+\mathbf{v}(t)\quad\mathbf{v}(t)\sim\mathcal{N}(\mathbf{0},\mathbf{R}(t))
\end{split}\,,
\label{eq:state_space_model}
\end{align}
where $\mathbf{Q}(t)$ and $\mathbf{R}(t)$ denote the covariance
matrices associated with the Gaussian process noise
$\mathcal{N}(\mathbf{0},\mathbf{Q}(t))$ and Gaussian observation noise
$\mathcal{N}(\mathbf{0},\mathbf{R}(t))$ at time $t$, respectively. We
assume the quantities $\mathbf{Q}(t)$ and $\mathbf{R}(t)$ to be
known. The function $\mathbf{f}(\cdot)$ describes the dynamics of the
system state $\mathbf{x}(t)$, while $\mathbf{h}(\cdot)$ maps
$\mathbf{x}(t)$ to a measurable quantity. Both functions can be
non-linear.
 \begin{figure*}
    \centering
    \includegraphics{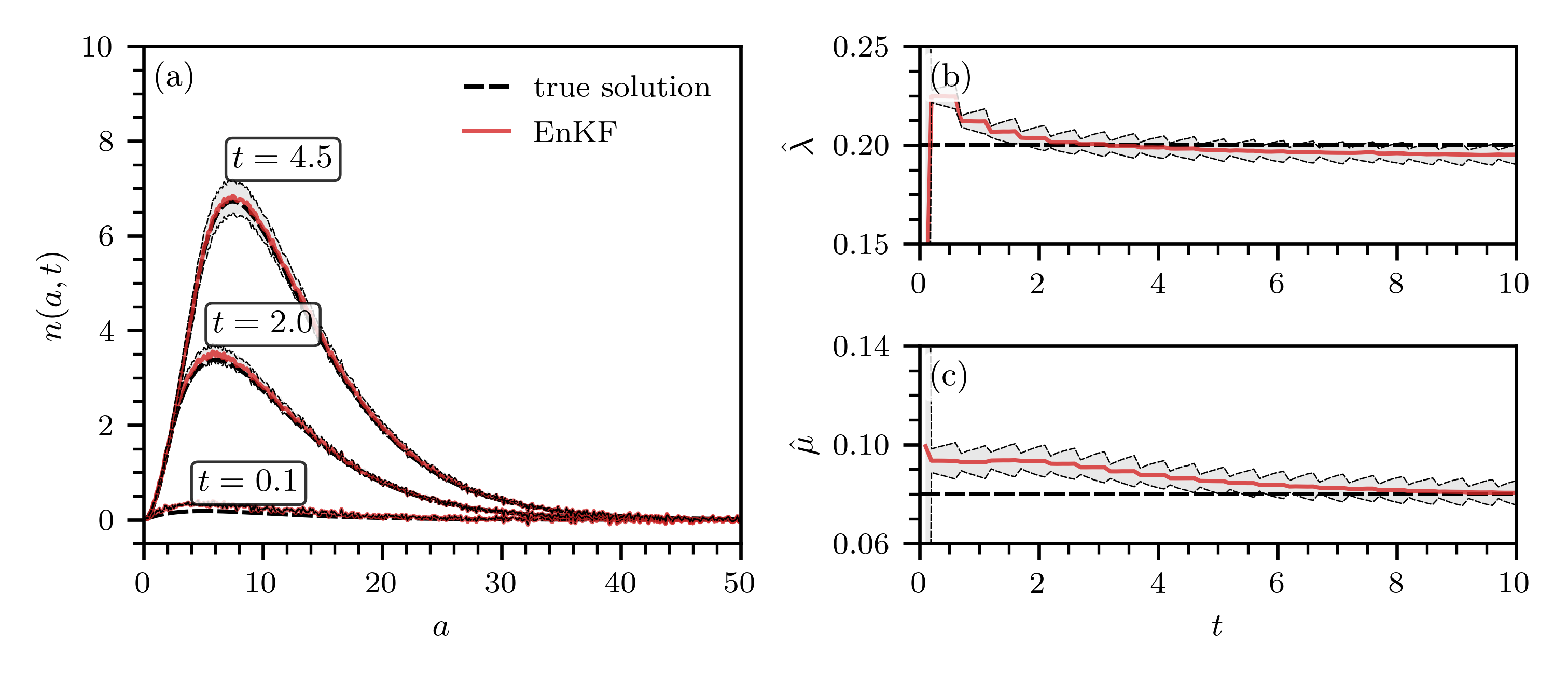}
    \caption
    {\textbf{State and parameter estimation with an EnKF.} (a) The
      population with SUD with age between $a$ and $a + \mathrm{d}a$
      at times $t=0.1,2.0,4.5$. (b,c) EnKF estimates
      $\hat{\lambda},\hat{\mu}$ of the rate parameters $\lambda,\mu$
      [see Eqs.~\eqref{back} and \eqref{front}]. In all panels, the
      true solution is represented by a dashed black line. Solid red
      lines and grey-shaded regions indicate EnKF solutions and
      corresponding $3\sigma$ intervals. The shown
      results are based on $M=500$ ensemble members.}
    \label{fig:da_simple_model}
\end{figure*}

In the context of the age-structured model \eqref{eq:pde_model},
element $x_j(t)$ of the state vector $\mathbf{x}(t)$ corresponds to
$n(a_j,t)\equiv n(a_0+j\Delta a,t)$ ($0\leq j\leq N_a-1$), the density
of individuals whose age lies within the $[a_0+j\Delta a,
  a_0+(j+1)\Delta a)$ interval at time $t$. That is,
\begin{equation}
\mathbf{x}(t)=[n(a_0,t),n(a_1,t),\dots]^\top\,.
\label{eq:unaugmented}
\end{equation}
We use $N_a$ and $\Delta a$ to denote the number of discretizations of
the age interval and the corresponding age discretization step,
respectively. For the numerical solution of
Eq.~\eqref{eq:state_space_model}, we later also discretize the
simulation time interval $[0,T]$ into $N_t$ equidistant intervals of
duration $\Delta t=T/N_t$. If we wish to estimate model parameters
such as $\mu$ and $\lambda$ introduced in Sec.~\ref{sec:model}, we can
augment the state to obtain
\begin{equation}
\mathbf{x}(t)=[n(a_0,t),\dots,n(a_{N_a-1},t),\mu,\lambda]^\top\,.
\label{eq:augmented}
\end{equation}
An example of an inference problem with an augmented state
\eqref{eq:augmented} will be provided in
Sec.~\ref{sec:example_estimation}.

At every time point $t$, the goal of filtering is to determine the
state posterior distribution given all prior observations. Before
producing EnKF state predictions, we generate an initial ensemble
$[\boldsymbol{\chi}_0^{(1)},\dots,\boldsymbol{\chi}_0^{(M)}]$ that
consists of $M$ ensemble members
$\boldsymbol{\chi}_0^{(i)}\sim\mathcal{N}(\hat{\mathbf{x}}_0,\mathbf{P}_0)$
($1\leq i \leq M$). The quantities $\hat{\mathbf{x}}_0$ and
$\mathbf{P}_0$ denote the given initial state and covariance
estimates, respectively.

We now outline the two main EnKF steps: (i) forecasting the evolution
of the system state and (ii) updating the predicted state estimates
using observation data. To do so, we discretize the time evolution of
the system state and use the shorthand notation $y_k\equiv y(t_k)$ to
refer to a quantity $y$ at time $t_k=k \Delta t$ ($0\leq k\leq
N_t$). Here and in the remainder of the manuscript, we assume that
$t_0=0$.

The basic idea behind forecast and update iterations is
  that one first uses state estimates $\boldsymbol{\chi}_{k}^{(i)}$ at
  time $t_k$ to calculate predicted state estimates
  $\boldsymbol{\chi}_{k+1}^{(i)-}$ at time $t_{k+1}$. These predicted
  estimates are the combined with observational data to obtain an
  updated state estimate $\boldsymbol{\chi}_{k+1}^{(i)}$. The superscript
  ``$-$'' in $\boldsymbol{\chi}_{k+1}^{(i)-}$ is used to distinguish
  the predicted (\ie, prior) state estimates from the updated (\ie,
  posterior) state estimates.

\begin{enumerate}
    \item[(i)] \textbf{Forecast Step:} For each ensemble member, we
      calculate the predicted state estimate
      $\boldsymbol{\chi}_{k+1}^{(i)-}$ according to
    \begin{equation}
    \boldsymbol{\chi}_{k+1}^{(i)-}=\boldsymbol{\chi}_{k}^{(i)}+\Delta t\,\mathbf{f}(\boldsymbol{\chi}_{k}^{(i)},t_k)+\boldsymbol{\epsilon}^{(i)}_{k}\,,    
    \label{eq:forecast_step}
    \end{equation}
where
$\boldsymbol{\epsilon}^{(i)}_{k}\sim\mathcal{N}(\mathbf{0},\mathbf{Q}_k)$.
For the sake of computational efficiency, we avoid
  discretizations of the partial derivative of $n(a,t)$ with respect
  to $a$ in the EnKF simulations. In all numerical experiments, we
  first derive closed-form expressions of the rate of change of
  $n(a,t)$ to compute predictions $\boldsymbol{\chi}_{k+1}^{(i)-}$
  according to Eq.~\eqref{eq:forecast_step}.
The ensemble mean of the predicted state, $\hat{\mathbf{x}}_{k+1}^-$,
and the corresponding covariance matrix,
$(\mathbf{P}_{\hat{\mathbf{x}} \hat{\mathbf{x}}}^-)_{k+1}$, are given
by
    \begin{widetext}
    \begin{align}
    \hat{\mathbf{x}}_{k+1}^-&=\frac{1}{M}\sum_{i=1}^M\boldsymbol{\chi}_{k+1}^{(i)-}\\
    (\mathbf{P}_{\hat{\mathbf{x}}\hat{\mathbf{x}}}^-)_{k+1}&=\frac{1}{M-1}\sum_{i=1}^M\left[\boldsymbol{\chi}_{k+1}^{(i)-}-\hat{\mathbf{x}}_{k+1}^-\right] \left[\boldsymbol{\chi}_{k+1}^{(i)-}-\hat{\mathbf{x}}_{k+1}^-\right]^\top\label{eq:P_xx}\,.
    \end{align}
    \end{widetext}
The covariance matrix
$(\mathbf{P}_{\hat{\mathbf{x}}\hat{\mathbf{x}}}^-)_{k+1}$ is not
required in the EnKF iteration, but it can be used to estimate
confidence intervals of $\hat{\mathbf{x}}_{k+1}^-$.

\item[(ii)] \textbf{Update Step:} We begin with deriving the ensemble
  mean of the predicted observation
    \begin{equation}
    \hat{\mathbf{z}}_{k+1}^-\equiv\frac{1}{M}\sum_{i=1}^M\mathbf{z}_{k+1}^{(i)-}=\frac{1}{M}\sum_{i=1}^M\mathbf{h}(\boldsymbol{\chi}_{k+1}^{(i)-})
    \end{equation}
as well as the corresponding covariances
    
    \begin{widetext}
    \begin{align}
    \begin{split}
      (\mathbf{P}_{\hat{\mathbf{z}}\hat{\mathbf{z}}}^-)_{k+1} & =\frac{1}{M-1}\sum_{i=1}^M
      \left[\mathbf{h}(\boldsymbol{\chi}_{k+1}^{(i)-})-\hat{\mathbf{z}}_{k+1}^- \right]
      \left[\mathbf{h}(\boldsymbol{\chi}_{k+1}^{(i)-})-
        \hat{\mathbf{z}}_{k+1}^-\right]^\top + \mathbf{R}_{k+1}\\
      (\mathbf{P}_{\hat{\mathbf{x}}\hat{\mathbf{z}}}^-)_{k+1} & =\frac{1}{M-1}\sum_{i=1}^M
      \left[\boldsymbol{\chi}_{k+1}^{(i)-}-\hat{\mathbf{x}}_{k+1}^-\right]
      \left[\mathbf{h}(\boldsymbol{\chi}_{k+1}^{(i)-})-\hat{\mathbf{z}}_{k+1}^-\right]^\top\,.
    \end{split}
    \end{align}
    \end{widetext}
    
    \noindent The Kalman gain is
    
    \begin{equation}
      \mathbf{K}_{k+1}=(\mathbf{P}_{\hat{\mathbf{x}}\hat{\mathbf{z}}}^-)_{k+1}
      (\mathbf{P}_{\hat{\mathbf{z}}\hat{\mathbf{z}}}^-)_{k+1}^{-1}\,.
    \end{equation}
For a given observation $\mathbf{z}_{k+1}$, the state update of
ensemble member $i$ is

    \begin{equation}
      \boldsymbol{\chi}_{k+1}^{(i)} = \boldsymbol{\chi}_{k+1}^{(i)-}+\mathbf{K}_{k+1}
      \left[\mathbf{z}_{k+1}+\boldsymbol{\eta}^{(i)}_{k+1}
        -\mathbf{h}(\boldsymbol{\chi}_{k+1}^{(i)-})\right]\,,
    \end{equation}
where
$\boldsymbol{\eta}^{(i)}_{k+1}\sim\mathcal{N}(\mathbf{0},\mathbf{R}_{k+1})$.
Finally, the updated state estimate and the corresponding covariance
matrix are given by

    \begin{align}
    \begin{split}
      \hat{\mathbf{x}}_{k+1}&=\frac{1}{M}\sum_{i=1}^M \boldsymbol{\chi}_{k+1}^{(i)}\\
    (\mathbf{P}_{\hat{\mathbf{x}}\hat{\mathbf{x}}})_{k+1} &
      =(\mathbf{P}_{\hat{\mathbf{x}}\hat{\mathbf{x}}}^-)_{k+1}-
      \mathbf{K}_{k+1}(\mathbf{P}_{\hat{\mathbf{z}}\hat{\mathbf{z}}}^-)_{k+1}\mathbf{K}_{k+1}^\top\,.  
    \end{split}
    \end{align}
\end{enumerate}
\subsection{Estimating model parameters}
\label{sec:example_estimation}
As a first example of estimating model parameters with the help of an
EnKF, we focus on the analytically solvable case from
Sec.~\ref{sec:model} for which closed-form analytical solutions of
$n(a,t)$ can be obtained. Our goal is to estimate $\mu$ and $\lambda$
in Eqs.\,\eqref{back} and \eqref{front}. We thus augment the state
\eqref{eq:unaugmented} by $\mu,\lambda$ to obtain
\begin{equation}
\mathbf{x}(t)=[n(a_0,t),\dots,n(a_{N_a-1},t),\mu,\lambda]^\top\,.
\label{eq:augmented_state_simple}
\end{equation}
In accordance with Eqs.~\eqref{back} and \eqref{front}, the evolution
of $n(a,t)$ is described by
\begin{align}
\begin{split}
\frac{\partial n(a,t)}{\partial t}=
\begin{cases}
(a-t)\, e^{-\lambda(a-t)-\mu  t}\quad &(a \geq t)\\
0\quad &(a<t)
\end{cases}
\end{split}\,.
\label{eq:simple_model_f}
\end{align}
The evolution of the first $N_a$ components of the augmented state
\eqref{eq:augmented_state_simple} is described by
      Eq.~\eqref{eq:simple_model_f}. We assume that we can observe
      perturbed versions of $n(a,t)$ but not $\mu,\lambda$ (\ie, the
      measurement function is
      $\mathbf{h}(\mathbf{x}(t))=[n(a_0,t),\dots,n(a_{N_a-1},t)]^\top$). To
      avoid sign changes in the estimates of $\mu,\lambda$ during the
      EnKF iterations, we apply an exponential transform to render
      both estimates positive. That is, we first replace
        $\mu,\lambda$ with $\tilde{\mu},\tilde{\lambda}$ in
        Eq.~\eqref{eq:augmented_state_simple} and then apply the
        transform
        $\mu=\exp(\tilde{\mu}),\lambda=\exp(\tilde{\lambda})$ before
        carrying out a prediction step according to
        Eq.~\eqref{eq:simple_model_f}.

In our simulations, we consider an age interval of $[0,120]$ years. We
set $N_a=1000$ such that $\Delta a=0.12~\mathrm{year}$, and we use a
timestep of $\Delta t=0.1~\mathrm{year}$. Process and observation
noise covariances are assumed to be time-independent and given by
$\mathbf{Q}=10^{-4}J_{N_a+2}$ and
$\mathbf{R}=\mathrm{diag}(10^{-4},\dots,10^{-4})$, respectively. Here,
$J_n$ denotes the $n\times n$ matrix of ones. Furthermore, we set the
initial state and its covariance matrix to
$\hat{\mathbf{x}}_0=[10^{-5},\dots,10^{-5},10^{-1},10^{-1}]$ and
${\mathbf{P}_0=\mathrm{diag}(0.5,\dots,0.5,1,1)}$, respectively.

We generate unperturbed observation data from the model using
$\mu=0.08/\mathrm{year}$ and $\lambda=0.2/\mathrm{year}$. The
perturbations that we add to $n(a,t)$ are normally distributed with
zero mean and variance $10^{-4}$. Our goal is, given the randomized
$n(a,t)$, to estimate the underlying $\mu$ and $\lambda$ with an EnKF
and verify the degree of accuracy of our estimates compared to the
original values. In real-world applications, new observation data may
not be available for each prediction. To account for this potential
lack of observation data, we perform update steps (\ie, integrate
observation data into our predictions) every five prediction periods.

Figure~\ref{fig:da_simple_model}(a) shows the evolution of both the
true solution $n(a,t)$ for which $\mu,\lambda$ are known (dashed black
lines) and of the corresponding EnKF estimates that use the augmented
state \eqref{eq:augmented_state_simple}. Grey-shaded regions indicate
3$\sigma$ intervals of the EnKF predictions [see Eq.~\eqref{eq:P_xx}].
We observe that the EnKF produces estimates of $\lambda$ and $\mu$
that are very close to the true solution after $t\gsim2$ years and
$t\gsim 7$ years, respectively.

\section{Application to drug-overdoses}\label{sec:application_wonder}
We now use an EnKF to combine the model in Eq.~\eqref{eq:pde_model}
with corresponding empirical data taken from the CDC WONDER
database. Here different causes of death are classified according to
the 10th revision of the International Statistical Classification of
Diseases and Related Health Problems (ICD-10). We selected ICD-10
codes T40 (poisoning by narcotics and psychodysleptics) and T43.6
(psychostimulants with abuse potential) and all drug-induced deaths,
including unintentional death, suicide, homicide, and death by an
undetermined cause. We extracted data for the period 1999 until 2020.

In order to interface drug-overdose data with the analytical setup
given in Eq.~\eqref{eq:pde_model}, we identify $n(a,t)\,\mathrm{d}a$
as the number of people with SUD (w.r.t.  any drug) of ages between
$a$ and $a +\mathrm{d}a$ at time $t$. We also associate the influx
into the SUD population with an addiction rate of the non-SUD
population: $r(a,t)[N(a,t) - n(a,t)]$, where $N(a,t)$ is the overall
population density at time $t$ from which we subtract $n(a,t)$, the
density of people with an existing SUD. Finally, the prefactor
$r(a,t)$ represents an age- and time-dependent addiction rate, which
might be modeled \cite{CHAOS} or estimated from additional data such
as surveys. Including these elements, the model in
Eq.~\eqref{eq:pde_model} is adapted to

\begin{align}
\begin{split}
 \left[\frac {\partial} {\partial a}  + \frac {\partial} {\partial t} \right] 
 n(a,  t)  =&   - \mu (a, t) n (a, t) \\
& + r(a,t) [N(a,t) - n(a,t)]\,.
\label{eq:pde_model2}
\end{split}
\end{align}
Equation~\eqref{eq:pde_model2} can be recast in the same form as
Eq. \eqref{eq:pde_model} via
\begin{equation}
\begin{aligned}
 \left[\frac {\partial} {\partial a}  + \frac {\partial} {\partial t} \right] 
 n(a,t)  = & - [\mu(a, t)  + r(a,t)]n(a, t) \\
\: &  \hspace{1.5cm} + r(a,t) N(a,t)\,.
\label{eq:pde_model3}
\end{aligned}
\end{equation}
Upon comparing Eq.\,\eqref{eq:pde_model3} to Eq.~\eqref{eq:pde_model}
we can identify $\mu(a,t) \to \mu(a,t) + r(a,t)$ and $p(a,t) \to
r(a,t)N(a,t)$, so that the analytical solutions to
Eq.~\eqref{eq:pde_model3} can be written through the proper
substitutions in Eqs.~\eqref{analysol1} and \eqref{analysol2}.  Apart
from $n(a,t)$, Eq.~\eqref{eq:pde_model3} contains the functions
$N(a,t), r(a,t), \mu(a,t)$. Here we introduce some possible forms for
them, based on available data and realistic assumptions. We begin with
the entire population density $N(a,t)$.
\begin{figure}
	\includegraphics{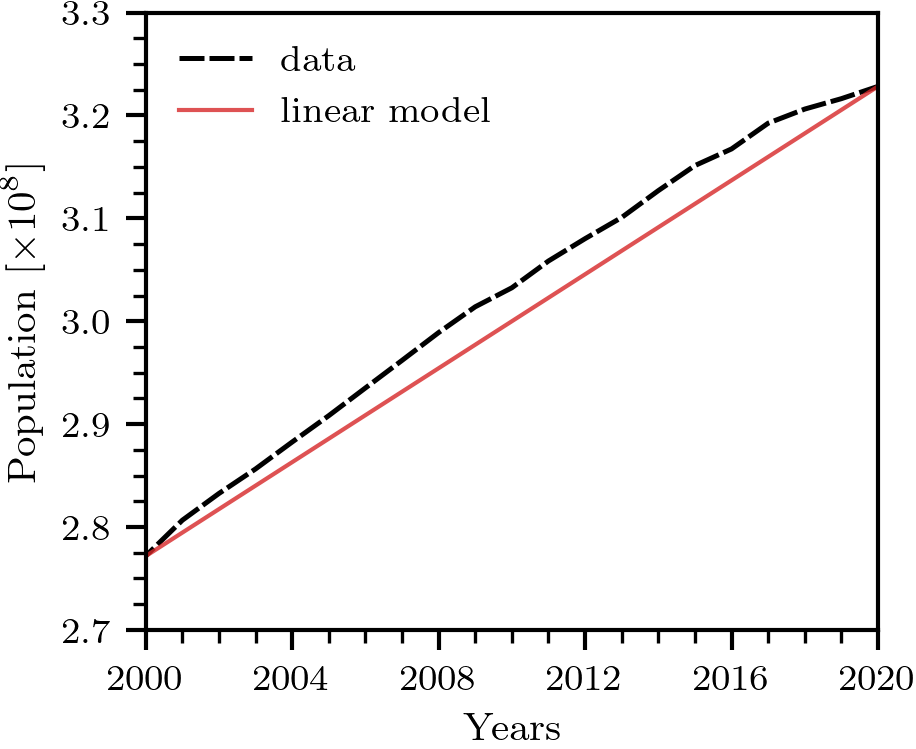}
	    \caption{\textbf{Population growth in the US
                  between 2000 and 2020.} The dashed black and solid
                red lines show the population growth in the US during
                2000--2020 and a linear
                population-growth model [see
                  Eq.~\eqref{eq:linear_population_growth}], respectively.}
    \label{fig:population_growth}
\end{figure}
\begin{figure*}
    \centering
    \includegraphics{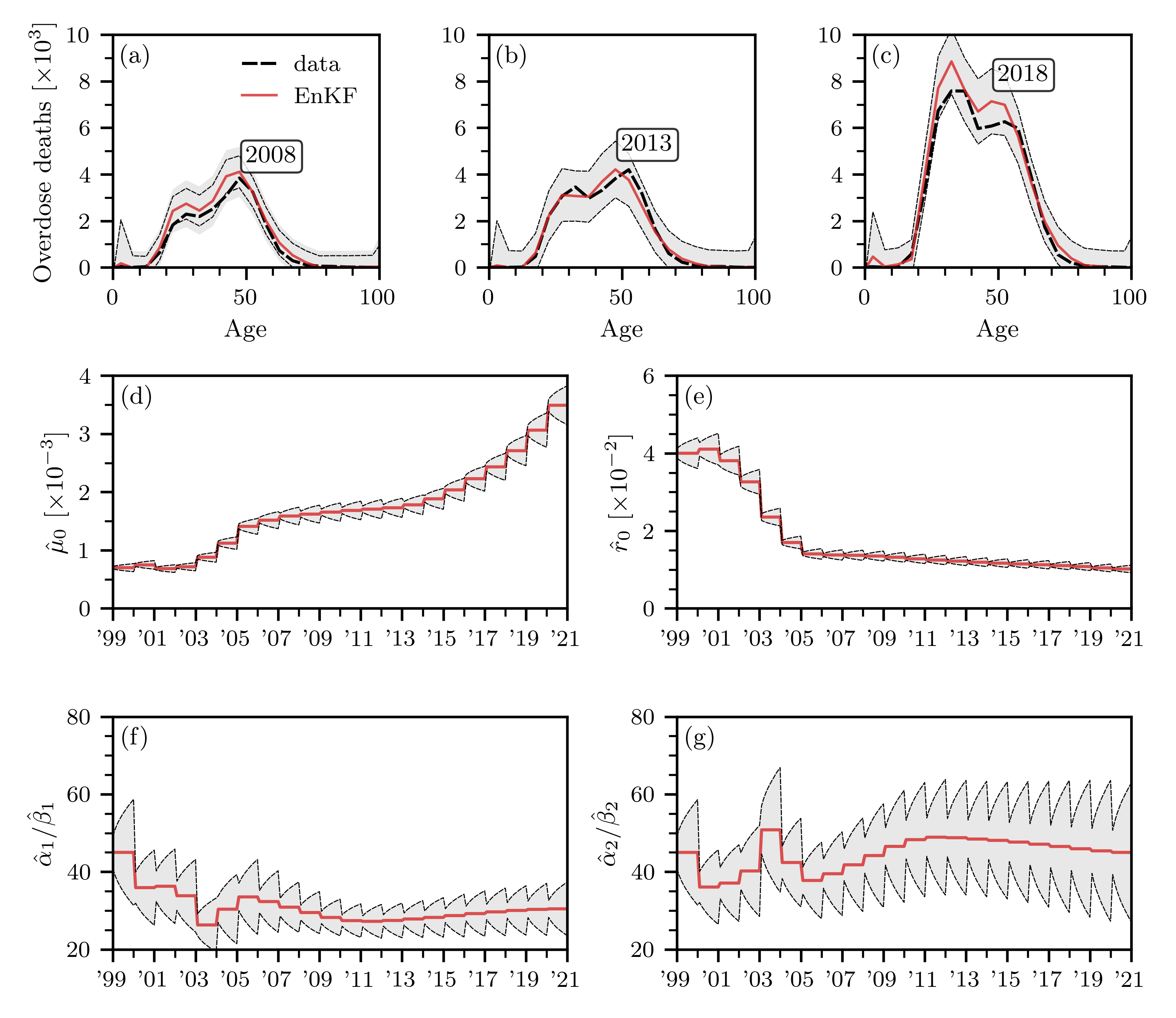}
    \caption{\textbf{Forecasting overdose mortality and estimating
        model parameters with an EnKF.} (a--c) Reported (dashed black
      lines) and predicted (solid red lines) numbers of overdose
      deaths in 2008, 2013, and 2018. Empirical data has been
      collected from the CDC WONDER database. (d--g) Evolution of
      estimated mortality rate $\hat{\mu}$, base modulating rate
      $\hat{r}_0$, and gamma function means
      $\hat{\alpha}_1/\hat{\beta}_1,\hat{\alpha}_2/\hat{\beta}_2$. In
      all panels, solid red lines and grey-shaded regions indicate
      EnKF solutions and corresponding $3\sigma$ intervals.  The shown
      results are based on $M=10^4$ ensemble members.  Observation
      data for the previous year becomes available in the beginning of
      every year.}
    \label{fig:crude_rates_data}
\end{figure*}
Because of different population-level dynamics such as aging and
immigration, the population growth in specific age classes is
non-monotonic. In principle, it is possible to use interpolation
methods and non-linear functions to construct an age-stratified
$N(a,t)$ based on empirical population data that is usually available
for 5 or 10-year age windows. However, for the sake of analytical
tractability, we will assume an age-independent quantity with $N(t)$
and focus on the more general form $N(a,t)$ in future work. To account
for the quasi-linear US population growth in the past two decades, we
set
\begin{equation}
N(t)=N_0+\Delta N t\,,
\label{eq:linear_population_growth}
\end{equation}
where $N_0=274.9\times 10^6$ and ${\Delta N=2.3\times
  10^6}$/year. Figure~\ref{fig:population_growth} shows
  the linear model $N(t)$ and the corresponding population data for
  the period 2000--2020.

To model $r(a,t)$, we assume a linear combination of two gamma
distributions $f_1(a;\alpha_1,\beta_1),f_2(a;\alpha_2,\beta_2)$, each
peaked at different ages, to describe possible variations in the age
dependence of the addiction rate. We do this to account for changes in
the prevalence of drug type and societal consumption patterns over the
21 year time frame we examine.  The quantities $\alpha_1,\beta_1$ and
$\alpha_2,\beta_2$ denote shape and rate parameters of the two
distributions $f_1$ and $f_2$, respectively. We also assume that
$r(a,t)$ does not depend on time and write
\begin{equation}
r(a,t)\equiv r(a)=\frac{r_0}{2}\left[f_1(a;\alpha_1,\beta_1)+f_2(a;\alpha_2,\beta_2)\right],
\end{equation}
where $r_0$ is a base modulating rate. 

The numerical results that we discuss in the following paragraphs show
that a linear combination of two gamma functions allows us to capture
the double-peaked distribution of age-stratified overdose deaths [see
  Fig.~\ref{fig:crude_rates_data}(a--c)].  Finally, for analytical
tractability of the double integrals arising from the solutions of
Eq.~\eqref{eq:pde_model3}, we retain the constant mortality rate
assumption $\mu(a,t)=\mu$. To combine the mechanistic model in
Eq.~\eqref{eq:pde_model3} with empirical data on overdose deaths, we
augment the system state $\mathbf{x}(t)$ [see
  Eq.~\eqref{eq:unaugmented}] by
\begin{equation}
\begin{split}
\tilde{D}(a_j,t)&=\int_0^t\mu(a_j,t')n(a_j,t')\,\mathrm{d}t'\\
& \hspace{3.6cm} (0\leq j\leq N_a-1)\,,
\end{split}
\end{equation}
where $\tilde{D}(a_j,t)$ is the cumulative number of overdose deaths
in the age interval $[a_j,a_{j+1})$ up to time $t$. We also augment
  the system state $\mathbf{x}(t)$ by the model parameters
  $\mu,r_0,\alpha_1,\beta_1,\alpha_2,\beta_2$ that we wish to
  estimate. As a result, the final augmented system state is
\begin{align}
\begin{split}
\mathbf{x}(t)=\Big[&n(a_0,t),\dots,n(a_{N_a-1},t),\\
& \hspace{6mm}\tilde{D}(a_0,t),\dots,\tilde{D}(a_{N_a-1},t),\\
& \hspace{14mm} \mu,r_0,\alpha_1,\beta_1,\alpha_2,\beta_2\Big]^{\top}.
\end{split}
\end{align}
We derive the corresponding rate of change
$\mathrm{\partial}n(a,t)/\mathrm{\partial}t$ for the EnKF updates in
Appendix~\ref{app1}.
\begin{figure*}
    \centering
    \includegraphics{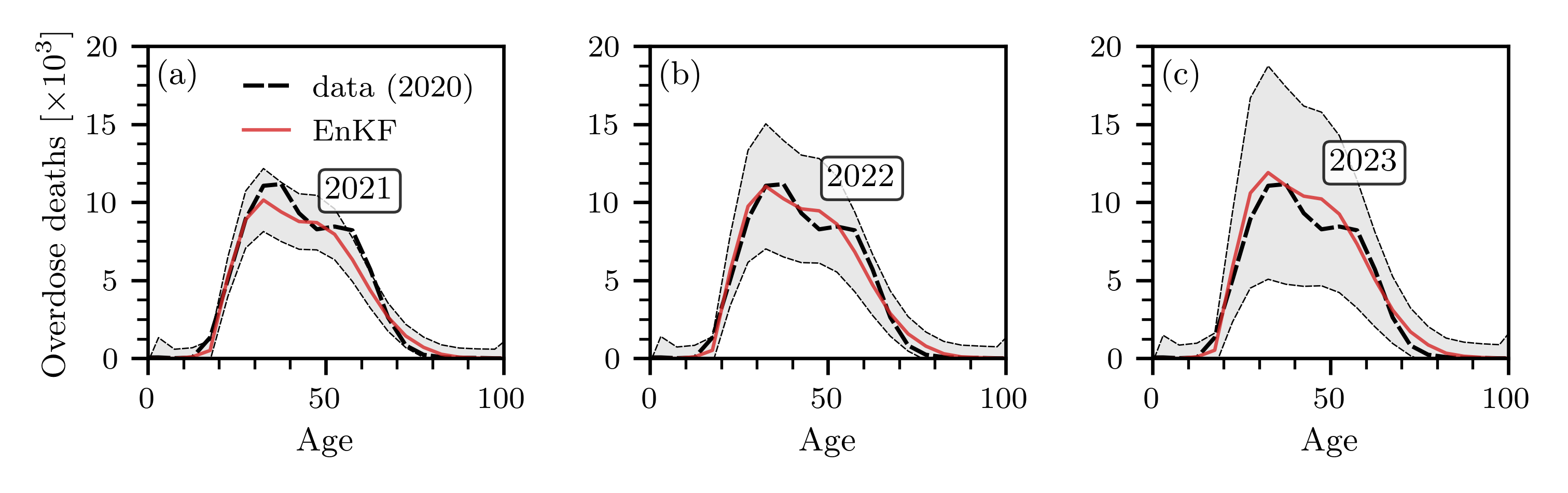}
    \caption{\textbf{Predicted overdose mortality in
          2021, 2022, and 2023.} Solid red lines show EnKF predictions
        of numbers of overdose deaths in (a) 2021, (b) 2022, and (c) 2023. As a
        reference, dashed black lines show overdose deaths in
        2020. Empirical data has been collected from the CDC WONDER
        database. Grey-shaded regions indicate corresponding $2\sigma$
        intervals. The shown results are based on $M=10^4$ ensemble
        members. The latest observation data that was available to
        generate the shown predictions is from 2020.}
    \label{fig:forecast}
\end{figure*}

For an accurate numerical evaluation of the rate of change of
$n(a,t)$, we use a sufficiently small discretization that is
associated with age windows that are more granular than the available
overdose data. We thus have to coarse-grain the modeled overdose death
densities to be able to relate them to observation data. The CDC
WONDER data that we use in this work is based on 22 age groups with
$a'_0=0,a'_{22}=120$ years and $\Delta a'_1=1,\Delta a'_2=4,\Delta
a'_3=5,\dots,\Delta a'_{21} = 5,\Delta a'_{22}=20$ years. We use a
superscript $'$ to distinguish the age discretization in the
observation data from the age discretization in the underlying model.

We combine the modeled quantities $\tilde{D}(a_j,t)$ with
corresponding observation data by numerically integrating
$\tilde{D}(a_j,t)$ over age windows $[a'_{\ell-1},a'_{\ell})$ ($1\leq
  \ell \leq 22$) to obtain the corresponding number of deaths
  $D(a'_\ell,t)$ in this age interval at time $t$. Here,
  $a'_\ell=a'_0+\sum_{m=1}^{\ell}\Delta a_{m}$ for $\ell\geq 1$. Based
  on the described mapping of $\tilde{D}(a_j,t)$ to $D(a'_\ell,t)$,
  the measurement function becomes
\begin{equation}
\mathbf{h}(\mathbf{x}(t))=[D(a'_1,t),D(a'_2,t),\dots]^\top\,.
\end{equation}
In our simulations, we set the initial values
$n(a_j,0)=\tilde{D}(a_j,0)=0$. The initial values of $\mu$, $r_0$,
$\alpha_1$, $\beta_1$, $\alpha_2$, $\beta_2$ are $7\times
10^{-4}\mathrm{/year}$, $0.04\mathrm{/year}$, $15$,
$1/(3~\mathrm{year})$, $15$, and $1/(3~\mathrm{year})$,
respectively. We have chosen the initial value of $r_0$ in accordance
with corresponding empirical data on the number of substance
initiates~\cite{abuse2019mental}. The initial mean of both gamma
distributions is equal to $\alpha_1/\beta_1=\alpha_2/\beta_2=45$
years. To ensure that the parameters $\mu$, $r_0$,
  $\alpha_1$, $\beta_1$, $\alpha_2$, $\beta_2$ stay positive during
  EnKF iterations, we use the same exponential transform as in
  Sec.~\ref{sec:example_estimation}. All initial covariances are set
to $10^{-4}$, except for the diagonal elements associated with $\mu$,
$r_0$, $\alpha_1$, $\beta_1$, $\alpha_2$, $\beta_2$, which are set to
$10^{-2}$. The process and observation noise covariances are as in
Sec.~\ref{sec:example_estimation}. We use a small process
  noise and a relatively large initial model parameter variance to (i)
  let the dynamics evolve according to the mechanistic drug-overdose
  model without too much additional noise in $n(a,t)$,
  $\tilde{D}(a,t)$ and (ii) let the filter explore different
  trajectories associated with appreciable variations in the
  underlying model parameters. We have also performed simulations for
  larger process noise values associated with $n(a,t)$ and
  $\tilde{D}(a,t)$. For example, we set the corresponding diagonal
  elements of $Q$ to values between 1 and 100 without observing
  substantial differences in the simulation results. In the
  measurement process, we divide $\tilde{D}(a,t)$ by $10^3$ to work
  with numerical values of $\mathcal{O}(1)$ when comparing predicted
  and observed overdose fatalities. A measurement variance of
  $10^{-4}$ (\ie, a standard deviation of $10^{-2}$) corresponds to
  about 10--100 overdose fatalities in the simulated data. Although
  the exact measurement noise is difficult to estimate given the
  unknown number of undocumented fatal drug-overdose cases, we used a
  standard deviation of about 10--100 as a reasonable modeling
  choice. In our simulations, we use a relatively large ensemble size
  of $M=10^4$ to minimize the effect of sampling errors that occur
  during the Monte Carlo approximation of the system state evolution
  in the prediction and update steps. Our simulations start in 1998
and we use a timestep of $\Delta t=0.1$ years.

Figure~\ref{fig:crude_rates_data}(a--c) shows reported drug-overdose
deaths (dashed black lines) for the years 2008, 2013, and 2018. Solid
red lines represent EnKF predictions that are based on updates that
involved observation data from all previous years since 1999. No
additional observation data were available between two subsequent
years. That is, for predictions that were made for, \eg, 2008, the
most recent observation data that was available to the EnKF was from
2007. Still, the EnKF predictions in
Fig.~\ref{fig:crude_rates_data}(a--c) are closely aligned with the
reported overdose deaths. For almost all age classes, predicted
overdose fatalities lie within the shown 3$\sigma$ regions (grey
shaded regions). For applications in real-time monitoring of overdose
fatalities, one may also include provisional data that becomes
available during the course of a year to further refine forecasts.

The evolution of $\hat{\mu}$ and $\hat{r}_0$ [see
  Fig.~\ref{fig:crude_rates_data}(d,e)] suggests that drug-overdose
mortality increased over the years while the proportion of newly
addicted individuals approaches about $0.01\mathrm{/year}$. Up until
2001-2002, the evolution of the mean values of both gamma functions is
synchronous [see Fig.~\ref{fig:crude_rates_data}(f,g)]. From 2003
onwards, one gamma function captures the addiction dynamics of
individuals who are older than 40 years, while the second gamma
function captures the inflow of younger people with substance use
disorder.

Finally, in Fig.~\ref{fig:forecast}, we show forecasts of
  overdose mortality in the US for the years 2021, 2022, and 2023. The
  latest observation data that is available for these forecasts is
  from 2020 (dashed black lines in Fig.~\ref{fig:forecast}). The
  predicted overdose mortality in 2021 is slightly smaller than in
  2020 in age groups between 30--60. In 2022 and 2023, the predicted
  overdose mortality in many age groups exceeds that of 2020. Since
  the overall overdose mortality has increased unsteadily more than
  five-fold in the past two decades (with a particularly steep
  increase between 2019 and 2020), the variance in the shown forecasts
  is relatively large.
\section{Discussion and Conclusions}\label{sec:discussion}
We have developed an age-structured model of drug-overdose
mortality. Our model accounts for age and time-dependent addiction and
mortality rates. It can readily be extended to account for multiple
drug classes and different ways of stratifying the population. In a
simple example, we have shown how age-structured models can be
combined with data-assimilation methods such as an ensemble Kalman
filter (EnKF) to forecast the evolution
of fatalities and estimate model parameters.

Combining our age-specific overdose model with empirical data on
overdose fatalities in the US, we have provided a proof-of-principle
set of methods that can be useful for estimating parameters governing
drug addiction and mortality and for forecasting the evolution of
population-level overdose dynamics.

In addition to developing a framework to include provisional overdose
data and retrospective updates of observation data, possible future
work includes the study of how regularization terms can help smooth
Kalman filter updates~\cite{JOH2008}, or how other ensemble-based
Kalman filters, such as ensemble adjustment Kalman
filters~\cite{anderson2001ensemble}, may help improve numerical
stability and forecast accuracy. For applications of the
  proposed methodology to small population sizes, it might be
  worthwhile to update the age-stratified population using a
  Poisson-process model, where the Gaussian noise term only affects
  the underlying model parameters and not the population numbers
  themselves. Since we focused on drug-overdose forecasting over the
  past two decades, we decided to use the EnKF in a forward mode and
  not use backward passes/smoothing (\ie, not use future observations
  from times $t'>t$ at time $t$). As noted by Evensen and van
  Leeuwen~\cite{evensen2000ensemble}, in forecasting mode, the EnKF
  and ensemble Kalman smoother (EnKS) produce the same state estimate
  at the latest time. However, using backward passes and an EnKS (or
  lagged versions) can help improve earlier parameter estimates, which
  is also an interesting direction for future research. Another
possible direction for future work is to extend the presented data
assimilation framework to account for age-dependent death rates
$\mu(a,t)$ and age-dependent population data $N(a,t)$ in
Eq.~\eqref{eq:pde_model3}.
\section*{Declarations}

\begin{itemize}
\item Funding: We acknowledge financial support from the ARO through
  grant W911NF-18-1-0345 (M.R.D and T.C.), the NIH through grant
  R01HL146552 (T.C.), and the NSF through grant DMS-1814090 (M.R.D.).

\item Conflict of interest/Competing interests: The authors declare no competing interests.
\item Availability of data and materials: All mortality datasets are publicly available at \url{https://gitlab.com/ComputationalScience/overdose-da} and at \href{https://wonder.cdc.gov/mcd.html}{https://wonder.cdc.gov/mcd.html}. 
\item Code availability: All source codes are publicly available at \url{https://gitlab.com/ComputationalScience/overdose-da}.
\item Authors' contributions: All authors contributed equally.
\end{itemize}

\noindent

\appendix
\section{Rate of change}
\label{app1}
We evaluate the derivative of Eq.~\eqref{analysol1} w.r.t.\ $t$ for
$t_0=0$, $r(a,t)\equiv r(a)=r_0
\left[f_1(a;\alpha_1,\beta_1)+f_2(a;\alpha_2,\beta_2)\right]/2$,
$N(t)=N_0+\Delta N t$, $\mu(a,t)=\mu$. The resulting rate of change of
$n(a,t)$ for $a>t$ is

\begin{widetext}
\begin{equation}
\begin{aligned}
\frac{\partial n(a,t)}{\partial t}= & -\big[\rho' (a -t)+\rho(a-t)(\mu+r(a-t))\big] e^{-\mu t-\int_{0}^t r(a-t+s)\, \mathrm{d}s}+r(a)N(t)\\
\: & \hspace{4mm}-\int_0^t e^{-\mu (t-s)-\int_{s}^t r(a-t+s')\, \mathrm{d}s'} N(s)\Big(r(a-t+s)\big(\mu+r(a-t+s)\big)+r'(a-t+s)\Big)\,\mathrm{d}s.
\end{aligned}
\end{equation}
For $a<t$, the rate of change is
\begin{align}
\frac{\partial n(a,t)}{\partial t}=\int_{0}^{a} \!r(s)\Delta N e^{-\mu(a-s)-\int_{s}^a r(s')\, \mathrm{d}s'}\, \mathrm{d}s\,.
\end{align}
The integrals $\int_{s}^t r(a-t+s')\, \mathrm{d}s'$, $\int_{0}^t
r(a-t+s)\, \mathrm{d}s$, and $\int_{s}^a r(s')\, \mathrm{d}s'$ can be
evaluated using the identity
\begin{equation}
\int_s^t \frac{\beta^\alpha}{\Gamma(\alpha)} (a-t+s')^{\alpha-1}e^{-\beta(a-t+s')}\,\mathrm{d}s'=\frac{1}{\Gamma(\alpha)}\left[\Gamma(\alpha,(a - t +s)\beta)-\Gamma(\alpha,a\beta)\right],
\end{equation}
\end{widetext}
\noindent
where $\Gamma(s,x)=\int_x^\infty t^{s-1}e^{-t}\,\mathrm{d}t$ denotes
the upper incomplete gamma function. We evaluate the remaining
integrals $\int_0^t (\cdot) \,\mathrm{d}s$ numerically. The simulation
results that we present in Sec.~\ref{sec:application_wonder} use the age-dependent initial condition
\begin{equation}
 \rho(a) = 0.04 N_0 f(a;\alpha,\beta)\,,
\end{equation}
where $f(a;\alpha,\beta)$ is a gamma distribution with shape and rate
parameters $\alpha$ and $\beta$, respectively. In all simulations, we
set $\alpha=15$ and $\beta=1/(3~\mathrm{year})$ such that the
distribution mean is 45 years. The prefactor of 0.04 is chosen such
that initially 4\% of the population are suffering from a substance
use disorder~\cite{abuse2019mental}.
\bibliography{sn-bibliography_2}

%
\end{document}